\documentclass[twocolumn,3p,times,procedia]{elsarticle}

\usepackage{ecrc}
\usepackage{graphicx}
\usepackage{balance}
\usepackage{array,threeparttable}
\usepackage{url}
 \usepackage{hyperref}

\volume{00}

\firstpage{1}

\runauth{}

\jid{procs}

\CopyrightLine{2022}{Published by Elsevier Ltd.}

\usepackage{amsmath,amsfonts,amssymb}
\usepackage{algorithm,algpseudocode}
\usepackage{varioref}
\usepackage{textcomp}
\usepackage{stfloats}
\usepackage{eqparbox}
\usepackage{verbatim}
\usepackage{subcaption} 
\usepackage{wrapfig}
\usepackage{enumitem}
\usepackage{listings}
\usepackage{ragged2e}
\usepackage{tikz}
\usepackage{xcolor}
\usepackage{mathrsfs}
\usepackage{empheq}
\usepackage{mathtools}
\usepackage[noabbrev]{cleveref}
\usepackage{booktabs}
\usepackage{longtable}
\usepackage[figuresright]{rotating}
\usepackage{tabularx}
\usepackage{bm}
\usepackage{fancyhdr}
\fancyheadoffset[RO,EL]{0pt}
\usepackage{geometry}
\begin{document}\small
\begin{frontmatter}

\dochead{}
\title{
\begin{flushleft}
{\LARGE Blockchain-Enhanced UAV Networks for Post-Disaster Communication: A Decentralized Flocking Approach} 
\end{flushleft}
}
 %

\author[]{ \leftline {Sana Hafeez, Runze Cheng, Lina Mohjazi, Yao Sun$^*$, Muhammad Ali Imran}}

\address{ \leftline {James Watt School of Engineering, University of Glasgow}
  \leftline{Glasgow G12 8QQ, UK}

}

\cortext[]{Corresponding Author: Yao Sun, Email: yao.sun@glasgow.ac.uk.}

\fntext[]{This paper was submitted in part to IEEE VTC2024-Spring, Singapore.}

\begin{abstract}
Unmanned Aerial Vehicles (UAVs) have significant potential for agile communication and relief coordination in post-disaster scenarios, particularly when ground infrastructure is compromised. However, efficiently coordinating and securing flocks of heterogeneous UAVs from different service providers poses significant challenges related to privacy, scalability, lightweight consensus protocols, and comprehensive cybersecurity mechanisms. This study introduces a robust blockchain-enabled framework designed to tackle these technical challenges through a combination of consensus protocols, smart contracts, and cryptographic techniques. First, we propose a consortium blockchain architecture that ensures secure and private multi-agency coordination by controlling access and safeguarding the privacy of sensitive data. Second, we develop an optimized hybrid consensus protocol that merges Delegated Proof of Stake and Practical Byzantine Fault Tolerance (DPOS-PBFT), aiming to achieve an effective balance between efficiency, security, and resilience against node failures. Finally, we introduce decentralized flocking algorithms that facilitate adaptable and autonomous operations among specialized UAV clusters, ensuring critical disaster relief functions under conditions of uncertain connectivity. Comprehensive simulations demonstrate the system achieved linear scaling of throughput up to 500 UAV nodes, with only a 50ms increase in latency from 10 to 500 nodes. The framework maintained high throughput and low latency despite spoofing, denial-of-service (DoS), and tampering attacks, showing strong cyber resilience. Communication latencies were kept under 10ms for diverse UAV operations through self-optimizing network intelligence, with median values around 2-3ms.
\end{abstract}

\end{frontmatter}


\section{Introduction}
\label{sec:intro}
Natural disasters, such as hurricanes, floods, and earthquakes, accidents can severely damage critical communication infrastructure, disrupting access to aid and relief coordination. Unmanned aerial vehicles (UAVs) offer the potential for rapidly restoring connectivity, but coordinating heterogeneous UAV flocks poses challenges in security, privacy, and scalability \cite{hafeez2023blockchain}. This motivates designing innovative blockchain-enabled UAV flocking networks to address limitations. This study proposes a novel framework using blockchain technology to improve UAV operations in post-disaster scenarios.
\subsection{Background}
Intelligent emergency communication systems are vital for ensuring effective network connectivity during disaster response scenarios. UAVs have proven effective at expanding wireless coverage for Internet-of-Things (IoT) devices due to their ability to hover in diverse locations and establish reliable links~\cite{derhab2023internet}. However, coordinating heterogeneous UAV fleets poses challenges such as limited flight endurance~\cite{keller2019natural}, restricted communication range~\cite{hafeez2023blockchain}, reliance on damaged ground networks and inadequate pre-planned routes~\cite{wang2023secure}, intermittent connectivity~\cite{damavsevivcius2023sensors}, lack of coordination between UAVs and human responders, security vulnerabilities from chaos, and insufficient transparency mechanisms.
Recent research has explored decentralized blockchain approaches to help overcome some obstacles through inherent attributes like distribution, security, transparency, automation, and resilience~\cite{cui2024efficient}. However, significant gaps remain before blockchain technology can be successfully incorporated into UAV networks for disaster response~\cite{ahad20246g}, despite analytical frameworks optimizing UAV deployment. Blockchain integration introduces new consensus, interoperability, security, and smart contract design challenges, specifically for decentralized disaster-resilient UAV fleet coordination~\cite{li2024formation}.

Although blockchain-enabled UAV networks have been studied to address security issues in UAV swarms~\cite{10051720}, research gaps remain in areas such as real-time data processing efficiency, scalability of blockchain solutions in large swarms, integration with existing air traffic control systems, and the development of standardized interoperability protocols among diverse UAV systems. Further research is also needed on the potential environmental impact and ethical considerations related to surveillance and data privacy.
Existing work focuses on building internal trust using blockchain~\cite{math11102262}, such as UAV practical Byzantine fault tolerance (U-PBFT) for lightweight consensus and real-time trust evaluation~\cite{zhang2023blockchain}. However, dynamic topology and limited UAV resources pose challenges~\cite{liu2024multistate}, highlighting the need for secure, efficient, and intelligent blockchain coordination frameworks tailored for disaster response UAVs to fully realize their decentralized collaborative autonomy potential.
This motivates designing a blockchain framework to realize the potential of UAVs for revolutionizing disaster response, addressing identified gaps through innovations in consensus protocols \cite{hafeez2024blockchainenabled}, interoperability, security mechanisms, and smart contract architectures specialized for resource-constrained UAV disaster response coordination.
\subsection{Motivation}
This study proposes several innovative solutions to address the identified limitations related to optimized blockchain consensus protocols, heterogeneous network interoperability, security against threats, and adaptive smart contracts for evolving disaster coordination needs. The overarching motivation is to transform disaster response strategies by unlocking the full potential of UAV networks through blockchain integration. Specifically, the aim is to facilitate decentralized, efficient, and autonomous UAV-based operations even in challenging post-disaster environments. This has the potential to significantly improve the effectiveness of time-critical relief efforts \cite{wang2024unmanned}. A hybrid Delegated Proof of Stake and Practical Byzantine Fault Tolerance (DPoS-PBFT) consensus approach that balances efficiency, security, and fault tolerance is promising for accommodating the constraints of the UAV platform and the volatility of the aerial environment. 
In addition, bio-inspired flocking techniques based on Reynolds rules \cite{olfati2006flocking} can enable resilient coordination of UAV clusters under uncertain connectivity during disaster relief. However, decentralized flocking alone cannot address critical security and access control challenges. Therefore, this study holistically tackles these motivations by developing an advanced decentralized ecosystem for secure, reliable, and optimized coordination among UAV fleets to enhance disaster management. The specific technical details that underpin the system design of the proposed novel framework are elaborated in Sections \ref{sec:model}- \ref{sec:hybrid_consensus}.

\subsection{Contributions and Organization}

\label{con}

In this study, we introduce a novel framework aimed at enhancing operations in wireless networks. Specifically, we propose strategies to optimize network performance within our framework, while also suggesting additional enhancements. Through simulations, we demonstrate the efficiency and adaptability of our proposed blockchain-based coordination framework, especially in dynamic environments with fluctuating resources, variable channel conditions, and diverse service requirements.
The key contributions of this study centre on pioneering advancements in system architecture, consensus protocols, and coordination algorithms to address existing limitations. The main contributions are
\begin{itemize}
    \item We propose a blockchain architecture enabling decentralized coordination across UAV networks, with a focus on preserving privacy and access control - crucial for efficient and secure disaster response.
    \item We design an optimized DPOS-PBFT consensus protocol for resource-constrained UAVs, balancing efficiency, security, fault tolerance and achieving lightweight processing, high throughput, low latency, and robustness.
    \item We present decentralized Reynolds flocking techniques enabling adaptable coordination of UAV swarms under uncertain connectivity.
    \item We demonstrate through comprehensive simulations that our proposed framework overcomes limitations in existing UAV network coordination. Specifically, we achieve excellent scalability, cyber resilience, and optimized latency profiles that unlock the potential of decentralized and intelligent UAVs for disaster response.
\end{itemize}
The rest of the paper is organized as follows. Section \ref{sec:lit_review} provides an overview of the relevant literature. Section \ref{sec:model} presents the architecture and the models of the proposed system. Section \ref{sec:Bio} details the decentralized flocking algorithm for UAV coordination. Section \ref{sec:hybrid_consensus} describes a customized hybrid consensus protocol. Section \ref{sec:results} presents an analysis of the simulation setup and the results. Finally, Section \ref{sec:conclusion} concludes the study and discusses future work.

\begin{table*}[ht!]
    \centering
    \caption{Comprehensive Overview of Challenges, Blockchain-Enabled Solutions, and Research Gaps in UAV Networks for Disaster Response.}
    \label{table:uav_networks}
    \begin{tabularx}{\textwidth}{>{\bfseries}lX}
        \toprule
        Category & Description \\
        \midrule
        \multicolumn{2}{c}{\textbf{Challenges in UAV-based Disaster Response \cite{keller2019natural}-\cite{maza2011experimental}}} \\
        \midrule
        Limited Flight Endurance & Battery dependence limits flight time and operational capability. \\
        Restricted Communication Range & Requires multi-hop routing which introduces delays. \\
        Reliance on Damaged Infrastructure & Reduces navigation and control effectiveness. \\
        Inadequate Pre-Planned Trajectories & Dynamic environments need real-time path replanning. \\
        Intermittent Connectivity & Weather or mobility lead to disruptions. \\
        Lack of Coordination & Between UAVs, ground robots, and human responders. \\
        Security Vulnerabilities & Spoofing, tampering, hijacking due to chaos. \\
        Communication and Delivery & Low efficiency in communication and relief delivery. \\
        Transparency and Accountability & Lack of transparency in relief management. \\
        Collaboration & Low collaboration due to different entity priorities. \\
        Computing and Energy & Centralized paradigms leading to failure points; energy constraints in UAV networks. \\
        Communication Channels & FL training issues due to UAV mobility and unreliable links. \\
        Privacy and Security & Data privacy concerns in UAV-based services. \\
        Caching and Delivery Services & Challenges in content caching and UAV deployment. \\
        \midrule
        \multicolumn{2}{c}{\textbf{Existing and Proposed Blockchain-Enabled Solutions \cite{hafeez2022beta}-\cite{hafeez2023birds}-\cite{hafeez2024blockchain}-\cite{hafeez2023blockchain}-\cite{mendis2020blockchain}}} \\
        \midrule
        Permissioned Blockchains & For efficiency, and privacy compared to public chains. \\
        Lightweight Consensus Protocols & E.g., PBFT, and PoA offer fault tolerance without mining. \\
        Smart Contracts & Automate coordination for flight plans, and information sharing. \\
        Tamper-Proof Data Logging & Using blockchain to record UAV data. \\
        Access Control via Smart Contracts & Secure coordination by authentication. \\
        Enhanced Data Integrity and Security & Secure and immutable ledger for data integrity. \\
        Improved Resource Allocation & Automated resource allocation via smart contracts. \\
        Decentralized Control & Reducing risks of central points of failure. \\
        Transparent Supply Chain Management & Real-time tracking and auditing of relief materials. \\
        Identity Management & Secure verification of individuals and organizations. \\
        Real-Time Data Sharing & Blockchain for efficient information exchange. \\
        Supply Chain Automation & Blockchain for logistics and supply chain optimization. \\
        Tokenization for Incentivization & Rewards for participation in disaster response. \\
        Interoperability Between Systems & Seamless data exchange and coordination. \\
        \midrule
        \multicolumn{2}{c}{\textbf{Open Research Challenges \cite{li2024formation}-\cite{wang2022platform}-\cite{duan2024design}}} \\
        \midrule
        Adaptive Coordination Algorithms & For dynamic environments and evolving needs. \\
        Handling Intermittent Connectivity & In UAV blockchain networks. \\
        UAV Computational Constraints & Limit complex chaincode and ledger size. \\
        Geo-spatial Smart Contract Support & Needed for location-based UAV coordination. \\
        Security Modeling and Analysis & Against threats like DDoS, and spoofing. \\
        Privacy Preservation & During UAV Surveillance Usage. \\
        Multi-Agency Collaboration & For large-scale disaster response. \\
        Optimization of UAV Payload and Range & For efficient delivery and operation. \\
        Robust Systems for Transparency & Ensuring accountability in relief operations. \\
        Multi-Stakeholder Collaboration Models & For effective information sharing. \\
        Sustainable Energy Solutions & For UAV Networks. \\
        Enhanced Privacy Protection Mechanisms & In UAV-based data gathering and AI model training. \\
        Advanced Algorithms for Content Caching & Efficient delivery services in UAV networks. \\
        Effective Tokenization Mechanisms & For incentivization in disaster response efforts. \\
        Standards and Protocols for Interoperability & Enhanced system integration in UAV networks. \\
        \bottomrule
    \end{tabularx}
\end{table*}
\section{Literature Review}
\label{sec:lit_review}
\subsection{Blockchain-Enabled UAV Solutions for Disaster Response}

UAV ad hoc networks have gained significant attention for their rapid deployment capabilities and resilience during disaster scenarios when ground infrastructure fails \cite{balakrishna2024device}. However, ensuring secure decentralized coordination within these dynamic networks poses several challenges. Blockchain technology offers a promising approach through its inherent features of distributed trust, transparency, and consensus mechanisms to address these challenges \cite{chughtai2024drone}. Nevertheless, consensus algorithms optimized for intermittent aerial links and resource-constrained UAV platforms are essential for effective implementation.

Existing blockchain solutions developed for vehicular networks have limited applicability when applied to the more volatile and dynamic nature of UAV swarms \cite{chughtai2024drone, venkatesan2024blockchain}. Numerous studies have explored blockchain-based systems for UAV coordination in disaster response scenarios, emphasizing secure information sharing \cite{raja2024ugen}, transparent data recording \cite{hafeez2023birds}, and accountability in relief distribution \cite{duan2024design}. However, challenges related to scalability, lightweight optimized consensus protocols for resource-constrained UAVs, and comprehensive privacy mechanisms remain largely unaddressed \cite{raja2024ugen}. Existing works in \cite{gc2023optimal} utilize smart contracts solely for agency authentication and access control, relying on centralized network components that limit robustness and resilience.

\subsection{Consensus Protocols and Smart Contracts for UAV Blockchains}

\begin{figure*}[htbp]
\centering
\includegraphics[width=0.7\linewidth]{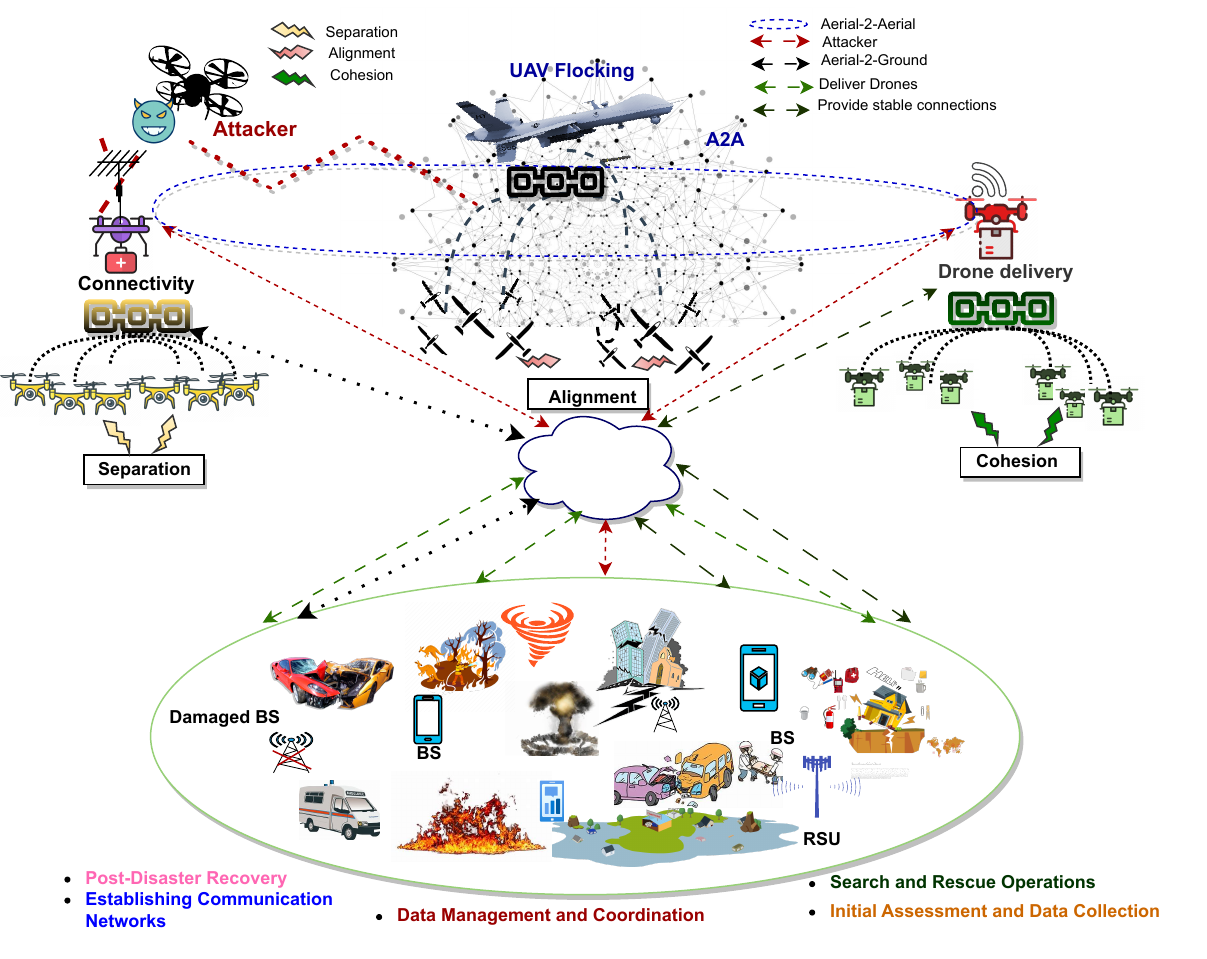}
\caption{The Architecture for Blockchain-Enabled UAV Coordination in Disaster Response.}
\label{fig:architecture}
\end{figure*}

The efficiency and fault tolerance of consensus protocols are critical factors for UAV networks, which typically involve resource-constrained nodes and intermittent connectivity. While the PBFT protocol provides resilience against byzantine failures, it suffers from high communication overhead \cite{sun2024joint}. On the other hand, the Proof-of-Authority (PoA) consensus mechanism reduces overhead by involving fewer validators but introduces centralization issues. To address the trade-offs between efficiency and security, recent research has proposed hybrid protocols that combine PBFT and PoA \cite{wang2024decentralized}. These hybrid approaches also incorporate techniques such as sharding and leveraging UAV mobility patterns to further improve throughput and reduce latency. Such optimized consensus protocols aim to achieve an effective balance between efficiency, security, and fault tolerance, making them suitable for time-sensitive UAV disaster response operations while maintaining the necessary security guarantees \cite{xing2022uavs}.

In the context of disaster management, smart contracts have been explored to automate coordination and ensure transparency by encoding rules that are executed based on predefined conditions. Proposed applications of smart contracts in UAV disaster networks include autonomous flight planning, decentralized information exchange between responders, and transparent tracking of aid distribution \cite{afotanwo2024exploring}. However, standard smart contract languages and data formats lack native support for spatial data required for geo-coordination of UAVs \cite{paulin2024application}. Novel geospatial smart contracts tailored for location-based UAV coordination show promise but require further research and development to address challenges such as efficient storage and querying of spatial data on the blockchain. Recent advancements in areas such as geospatial smart contracts \cite{xing2022uavs}, disaster-resilient communication protocols \cite{iyer2023perspectives}, and privacy-preserving UAV coordination techniques \cite{nicolazzo2024privacy} show promise in overcoming these limitations. This motivates the design of a comprehensive framework that emphasizes decentralization, efficiency, privacy, and resilience, tailored specifically for secure, blockchain-enabled coordination in real-world UAV-assisted disaster response scenarios. Table \ref{table:uav_networks} summarizes the key challenges, existing solutions, and open research problems in blockchain-enabled UAV networks for disaster response scenarios. 

In summary, UAV ad hoc networks are increasingly recognized for their rapid deployment capabilities and resilience when ground infrastructure is compromised \cite{mohsan2023unmanned}. However, secure decentralized coordination within these dynamic networks remains a significant challenge. While blockchain technology offers distributed trust and consensus mechanisms to address these challenges, algorithms specifically tailored for intermittent aerial communication links and resource-constrained UAV platforms are essential \cite{gc2023optimal}-\cite{mohsan2023unmanned}.
Existing blockchain solutions for vehicular networks have limited applicability to the more volatile nature of UAV swarms \cite{hadi2023comprehensive}. Although UAV consensus protocols have been proposed, they often lack considerations for fluctuating nodes, link conditions, and the unique constraints of UAV platforms \cite{javed2024state}. Numerous studies have focused on blockchain systems for disaster-response UAV coordination, addressing secure information sharing, transparent data recording, and accountability in relief distribution \cite{bibri2024smarter}. However, issues remain with scalability, lightweight consensus protocols for resource-constrained UAVs, comprehensive privacy mechanisms, and effective utilization of smart contracts for spatial coordination of UAV fleets. Recent advancements in areas such as geospatial smart contracts, resilient communication protocols, and privacy-preserving coordination techniques show promise in addressing these limitations \cite{kirli2022smart}. This study aims to advance secure, blockchain-enabled coordination specifically tailored for real-world disaster-response UAVs, with a focus on decentralization, efficiency, privacy, resilience, and effective utilization of smart contracts for spatial coordination of heterogeneous UAV fleets.

This Fig. \ref{fig:architecture} illustrates the deployment and coordination of UAVs in a post-disaster scenario. It depicts various UAV operations and activities aimed at facilitating effective disaster response and recovery efforts. The image shows UAV flocks engaged in different tasks, such as search and rescue operations, initial assessment and data collection, data management and coordination, and post-disaster recovery efforts. The UAVs are organized into flocks and exhibit flocking behaviour, characterized by separation, cohesion, and alignment, which enables coordinated movement and efficient coverage of the affected areas.
The figure also highlights the establishment of communication networks, including aerial-to-aerial (A2A) and aerial-to-ground (A2G) links, to provide stable connectivity and enable drone delivery services. These communication networks are crucial for maintaining reliable communication channels, coordinating UAV operations, and facilitating the delivery of essential supplies to affected regions.

Additionally, the diagram depicts an attacker's presence, indicating the potential security threats and the need for robust cybersecurity measures to protect the UAV networks and their operations from malicious actors. The diagram also shows damaged base stations (BS) and a roadside unit (RSU), representing the disruption of ground infrastructure commonly experienced in disaster scenarios. This emphasizes the importance of UAVs in providing alternative means of communication and support when traditional infrastructure is compromised. Overall, this main scenario illustrates the various components and activities involved in a coordinated UAV-based disaster response effort, highlighting the importance of communication networks, flocking behaviour, drone delivery services, and the need for security measures to ensure effective and resilient operations in challenging post-disaster environments.
\section{System Architecture and Models}
\label{sec:model}
This section describes the mathematical models used to characterize the architecture for UAV-based disaster response. Specifically, it covers models related to communication, mobility, flocking algorithms, reliability, and security. The communication model under discussion is primarily centred around the propagation characteristics of wireless links between UAVs in an aerial network. A pivotal element of this model is the \textit{log-distance path-loss model}. This model is prevalently used for modelling signal attenuation in A2A channels, especially in scenarios involving UAVs. The \textit{log-distance path-loss model} is a fundamental concept in wireless communications. It is utilized to estimate the loss of signal strength, known as path loss, over a distance. The model calculates this loss based on the logarithm of the distance between the transmitter and the receiver. It takes into account the path-loss exponent and the loss at a reference distance, making it particularly relevant in the context of UAV communications. This relevance stems from its utility in understanding and predicting signal strength variations over different distances in three-dimensional airspace.
\subsection{Communication Model}
\label{sec:comm_model}
The communication model defines the propagation characteristics of the wireless links between UAVs in an aerial network. A2A propagation relies on the log-distance path-loss model, which is commonly used for modelling signal attenuation in A2A channels.
The path loss $\mathrm{PL}(d_{ij})$ depends on the distance $d_{ij}$ between the transmitting UAV $i$ and receiving UAV $j$. It describes how the signal strength decays with distance as it propagates through the medium. The path loss is calculated as
\begin{equation}
\mathrm{PL}(d_{ij}) = \mathrm{PL}_0 + 10n \log_{10}\left(\frac{d_{ij}}{d_0}\right).
\end{equation}
where $n$ is the path-loss exponent, $d_0$ is the reference distance, and $\mathrm{PL}_0$ is the path loss at the reference distance $d_0$. The path-loss exponent $n$ depends on the specific environment. 
Using this path loss model, the received signal-to-noise ratio (SNR) between UAVs $i$ and $j$ can be computed as
\begin{equation}
\mathrm{SNR}_{ij} = P_t + \mathcal{G}_i + \mathcal{G}_j - \mathrm{PL}(d_{ij})
\label{eq:snr}
\end{equation}
where $P_t$ is the transmit power of UAV $i$, and $\mathcal{G}_i$ and $\mathcal{G}_j$ are the antenna gains of UAVs $i$ and $j$, respectively. The 3D positions $\mathbf{P}_i$ and $\mathbf{P}_j$ determine the separation distance $d_{ij}$, which directly affects the path loss.
The maximum achievable data rate $R_{ij}$ for the A2A link is calculated using Shannon's capacity formula
\begin{equation}
R_{ij} = B\log_{2}(1 + \mathrm{SNR}_{ij}).
\label{eq:rate}
\end{equation}
where $B$ denotes the channel bandwidth, sustaining adequately high mesh link data rates is crucial for reliable UAV coordination and message exchange control. To model mobility, 3rd Generation Partnership Project (3GPP) TR 36.777 is referenced, which provides standard statistical 3D trajectory models.
 \footnote{\url{https://www.3gpp.org/ftp/Specs/archive/36_series/36.777/}}
This allows the capturing of realistic fluctuations in UAV trajectories.
\subsection{Consortium Blockchain Architecture}
This section outlines the consortium blockchain architecture tailored for decentralized coordination and access control in disaster response scenarios, involving both service providers and UAV networks. We focus on key parameters such as transaction throughput $T(N)$ and latency $L(N)$, which are crucial for assessing the system's performance and are analyzed using mathematical methods.
The architecture is specifically designed to enhance communication, coordination, and data sharing among the involved parties. Utilizing smart contracts establishes robust access control policies. Each participating entity is assigned specific permissions $P(x)$, dictating their level of access within the network. The function of the smart contract, denoted as $\mathrm{SC_{ac}}(P(x), ac_i) \rightarrow \{\text{Allow, Deny}\}$, is to enforce these policies based on a set of predefined rules.

Performance metrics such as $T(N)$ and $L(N)$, where $N$ represents the number of nodes in the network, are analyzed to understand their impact on the system's overall efficiency. This analysis is particularly important as it provides insights into how the system's performance might vary with changes in scale and node density during disaster response operations. Moreover, the use of cryptographic methods, including zero-knowledge proofs, is highlighted as a means to facilitate privacy-preserving coordination, especially when handling sensitive information.

\subsection{Security Measurements}
Analytical models are used to quantify the overall risk $\mathcal{X}(t)$ and resilience $R(t)$ based on various threat factors, including denial-of-service, spoofing, and tampering. This enables the evaluation of security mechanisms.
We modelled the security risks faced by the UAV network, such as denial-of-service attacks, spoofing, tampering, and malware infections. The overall risk $\mathcal{X}(t)$ at time $t$ is given by
\begin{equation}
\mathcal{X}(t) = w_{\mathcal{D}} D(t) + w_{\mathcal{S}} S(t) + w_{\mathcal{T}} T(t) + w_{\mathcal{M}} M(t),
\end{equation}
where $D(t)$, $S(t)$, $T(t)$, and $M(t)$ represent the individual risk factors, and $w_{\mathcal{D}}$, $w_{\mathcal{S}}$, $w_{\mathcal{T}}$, and $w_{\mathcal{M}}$ are the weights for tuning their relative importance. The individual risk factors are modelled as follows
\begin{align}
D(t) &= \lambda_{\mathcal{D}} e^{-\mu_{\mathcal{D}} t}, \\
S(t) &= \lambda_{\mathcal{S}} (1 - e^{-\mu_{\mathcal{S}} t}), \\  
T(t) &= \lambda_{\mathcal{T}} t e^{-\mu_{\mathcal{T}} t}, \\
M(t) &= \lambda_{\mathcal{M}} (1 - e^{-\mu_{\mathcal{M}} t}).
\end{align}
Here, $\lambda$ denotes the initial risk magnitude, and $\mu$ represents the mitigation rate for each threat. This allows for an analytical quantification of the evolution of risk.
The resilience $R(t)$ is measured as
\begin{equation}
R(t) = 1 - \frac{\mathcal{X}(t)}{\mathcal{X}(0)},
\end{equation}
where $\mathcal{X}(0)$ denotes the initial risk. Simulations assess $R(t)$ under different attack scenarios and intensities. The results demonstrate that the system maintains resilience, with $R(t) > 80\%$ under realistic threat levels, owing to the implemented security mechanisms. Quantitative evaluation of resilience through analytical modelling assures the robustness of the system against evolving attacks.

\begin{figure*}[ht]
\centering
\includegraphics[width=0.7\linewidth]{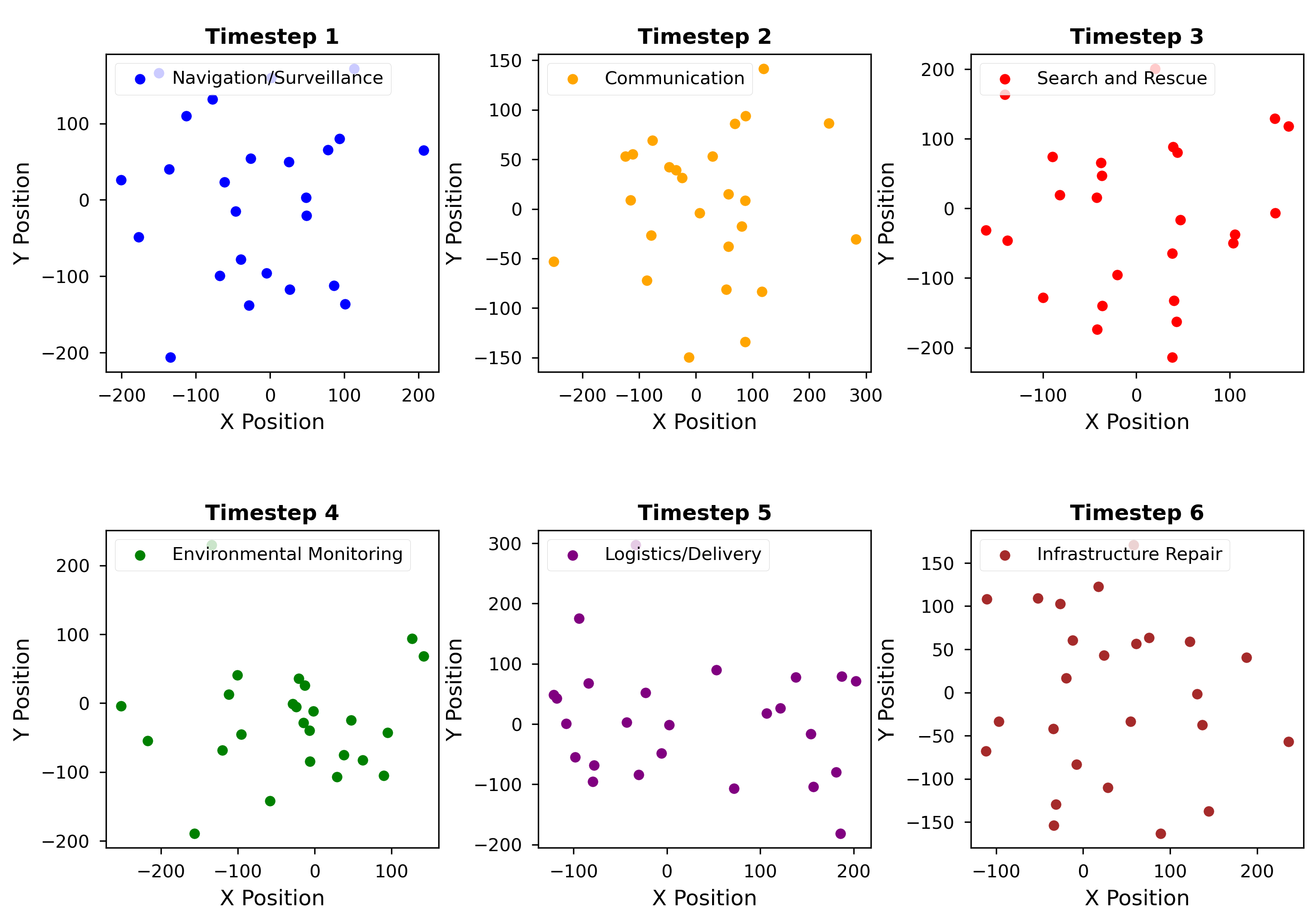}
\caption{The 2D spatial distribution of flocking UAVs engaged in post-disaster activities. More detailed description is provided in Section~\ref{sec:Bio}.}
\label{fig:flocking}
\end{figure*}

\section{Decentralized Flocking Model for UAV Disaster Response}
\label{sec:Bio}

This section presents a decentralized flocking model designed to enable resilient coordination among specialized UAV clusters conducting critical disaster-relief functions, even amidst disrupted connectivity.

\subsection{Concrete Examples of Flocking Algorithms for UAV Disaster Relief Functions}
The proposed UAV network is heterogeneous and comprises several specialized clusters: the delivery network $S_{\lambda}$, focused on transporting relief supplies; the survey network $S_{\eta}$, assigned to rapid damage assessment; and the connectivity network $S_{\Omega}$, responsible for restoring communication links. A central UAV monitor, $\Upsilon_m$, dynamically adjusts high-level coordination strategies based on evolving disaster-response priorities. The delivery flock $\mathcal{S}_{\lambda}$ plays a pivotal role in immediate relief efforts by transporting essential supplies to affected areas. Using flocking algorithms, these UAVs maintain cohesion, alignment, and separation to ensure efficient and safe delivery of aid. The survey flock $\mathcal{S}_{\eta}$ focuses on damage assessment and mapping. Employing flocking strategies, these UAVs can systematically cover disaster areas, maintain communication, and avoid collisions. The connectivity flocks $\mathcal{S}_{\Omega}$ are aligned with 3GPP UAV standards to ensure efficient communication restoration in disaster-stricken areas. UAVs use flocking rules to maintain optimal formation for wireless coverage and to navigate safely through the environment.

The central monitor UAV, $\Upsilon_m$, coordinates the activities of these flocks by utilizing a decentralized coordination algorithm based on Reynolds flocking rules. The control input $\varphi_i$ for each UAV $\Upsilon_i$ comprises terms for separation, alignment, cohesion, and navigation, allowing for collision avoidance and coordinated trajectory planning. Additionally, a dynamic dissipating obstacle avoidance mechanism is incorporated, enabling UAVs to effectively navigate around obstacles.

\subsection{Reynolds Flocking Rules and Their Application}
The Reynolds flocking rules guide the decentralized coordination of the UAVs in our model. These rules consist of four key components: separation ($\varphi^s_i$), alignment ($\varphi^a_i$), cohesion ($\varphi^c_i$), and navigation ($\varphi^n_i$). Each UAV $\Upsilon_i$ receives a control input $\varphi_i$, which is a combination of these components that facilitates coordinated movement while avoiding collisions.

Obstacle Avoidance Dynamics: To navigate effectively around obstacles, UAVs employ dynamically dissipating obstacle avoidance. This dynamic is mathematically represented as
\begin{equation}
\langle \mathbf{v}_{\gamma}, \bar{\mathbf{v}} \rangle \geq \cos(\vartheta_m) |\bar{\mathbf{v}}|^2.
\end{equation}
Here, $\vartheta_m$ denotes the maximum allowable misalignment angle between the UAV's velocity vector $\mathbf{v}_{\gamma}$ and the desired direction $\bar{\mathbf{v}}$.
In our model, each UAV is considered an autonomous agent with a state comprising its position and velocity vectors, denoted by $(x_i, v_i)$ for UAV $i$. The control protocol for obstacle avoidance, essential in cluttered post-disaster environments, is the sum of three terms $u_i = u_i^{\alpha} + u_i^{\beta} + u_i^{\gamma}$, where each term represents specific control aspects for the UAV.
Our approach emphasizes a peer-to-peer control mechanism, eschewing a centralized command structure. This design enhances the system's resilience and adaptability. A central monitoring UAV, $\Upsilon_m$, orchestrates the overall flocking behaviour and adapts to dynamic disaster response needs. The primary objective of this decentralized flocking system is to enable robust and autonomous coordination among UAV flocks, thereby facilitating key disaster response tasks. Each UAV's state is defined by its position $\rho_i$, velocity $\Omega_i$, and designated flock type $\zeta_i$, ensuring that every UAV contributes optimally to the overall mission.
\subsection{Dynamic State Propagation and Battery Model}
The state of each UAV evolves according to
\begin{align}
    \rho_i[\kappa+1] &= \rho_i[\kappa] + \Delta \tau \cdot \Omega_i[\kappa], \\
    \Omega_i[\kappa+1] &= \Omega_i[\kappa] + \Delta \tau \cdot (\varphi^s_i + \varphi^a_i + \varphi^c_i + \varphi^n_i).
\end{align}
Here, $\Delta \tau$ represents the discrete time step. The battery dynamics of the UAVs are modelled to account for power.
This Fig. \ref{fig:flocking} presents the 2D spatial distribution of flocking UAVs engaged in post-disaster activities, including Navigation/Surveillance, Communication, Search and Rescue, Environmental Monitoring, Logistics/Delivery, and Infrastructure Repair, across consecutive timesteps labelled 1 through 6. Each scatter plot represents a specific timestep, with the x and y axes indicating the geographical coordinates within a 2D plane. The scattered dots within each plot represent individual UAVs, with their positions depicting the flocking patterns, deployment areas, and coverage for the corresponding post-disaster activity during that particular time interval.

\subsection{Significance of Flocking Algorithms in Multi-Agent Systems}
Consider a group of autonomous agents $\mathcal{A} = \{A_{1}, A_{2}, \ldots, A_{N}\}$, where each agent $A_{i}$ has a state $(x_{i}, v_{i}) \in \mathbb{R}^{n} \times \mathbb{R}^{n}$ representing its position and velocity vectors, respectively.
The control input $U_{i}$ for each agent $A_{i}$ comprises three terms
\begin{equation}
U_{i} = \hat{U}_{i}^{\text{coh}} + \bar{U}_{i}^{\text{damp}} + \check{U}_{i}^{\text{nav}},
\end{equation}
where $\hat{U}_{i}^{\text{coh}}$ enables cohesion towards the flock center, $\bar{U}_{i}^{\text{damp}}$ achieves velocity consensus through damping force, and $\check{U}_{i}^{\text{nav}}$ drives navigation towards the group objective.
We propose two flocking algorithms based on different interaction rules
\begin{equation}
U_{i} = U_{i}^{\alpha},
\end{equation}
where,
\begin{equation}
U_{i}^{\alpha} = \underbrace{\sum_{A_{j} \in \mathcal{N}_{i}} \phi_{b}(\lVert x_{j} - x_{i} \rVert_{\sigma}) \mathbf{n}_{ij}}_{\text{Cohesion Term}} + \underbrace{\sum_{A_{j} \in \mathcal{N}_{i}} a_{ij}(x) (v_{j} - v_{i})}_{\text{Damping Term}},
\end{equation}
where $\phi_{b}$ is used for the Reynolds flocking rule terms, and $\mathcal{N}_{i}$ are the neighbor sets for agent $i$.
\subsection{Alpha-Neighbors of Alpha-Agents: Proximity Net}
Let $\mathcal{V}_{\alpha} = \{1, 2, \ldots, n_{\alpha}\}$ and $\mathcal{V}_{\beta} = \{1, 2, \ldots, n_{\beta}\}$ denote the sets of alpha and beta agents, respectively. An obstacle $\beta_k \in \mathcal{V}_{\beta}$ is a neighbor of alpha-agent $i \in \mathcal{V}_{\alpha}$ if
\begin{equation}
\mathcal{B}(q_i, r_{\beta}) \cap O_k \neq \emptyset,
\end{equation}
where $\mathcal{B}(q_i, r_{\beta})$ is a ball centered at $q_i$ with radius $r_{\beta}$, and $O_k$ is the obstacle region. The alpha and beta neighbour sets are defined as
\begin{align}
\mathcal{N}_{\alpha i} &= \{j \in \mathcal{V}_{\alpha} : \|q_j - q_i\| < r_{\alpha}\}, \\  
\mathcal{N}_{\beta i} &= \{k \in \mathcal{V}_{\beta} : \mathcal{B}(q_i, r_{\beta}) \cap O_k \neq \emptyset\}.
\end{align}
where $q_i$ and $v_i$ are the position and velocity of agent $i$ in the obstacle boundary dynamics, respectively. This induces a bipartite proximity graph $\mathcal{G} = (\mathcal{V}, \mathcal{E})$ between the alpha and beta agents, where $\mathcal{V} = \mathcal{V}_{\alpha} \cup \mathcal{V}_{\beta}$ and $\mathcal{E} \subseteq \mathcal{V}_{\alpha} \times \mathcal{V}_{\beta}$. Here, $r_{\alpha}$ and $r_{\beta}$ are the radii for proximity in the alpha and beta neighbour sets, respectively.
\section{Enhanced DPoS-PBFT Consensus Mechanism for UAV Networks}
\label{sec:hybrid_consensus}
\begin{figure*}
\centering
\includegraphics[width=0.7\linewidth]{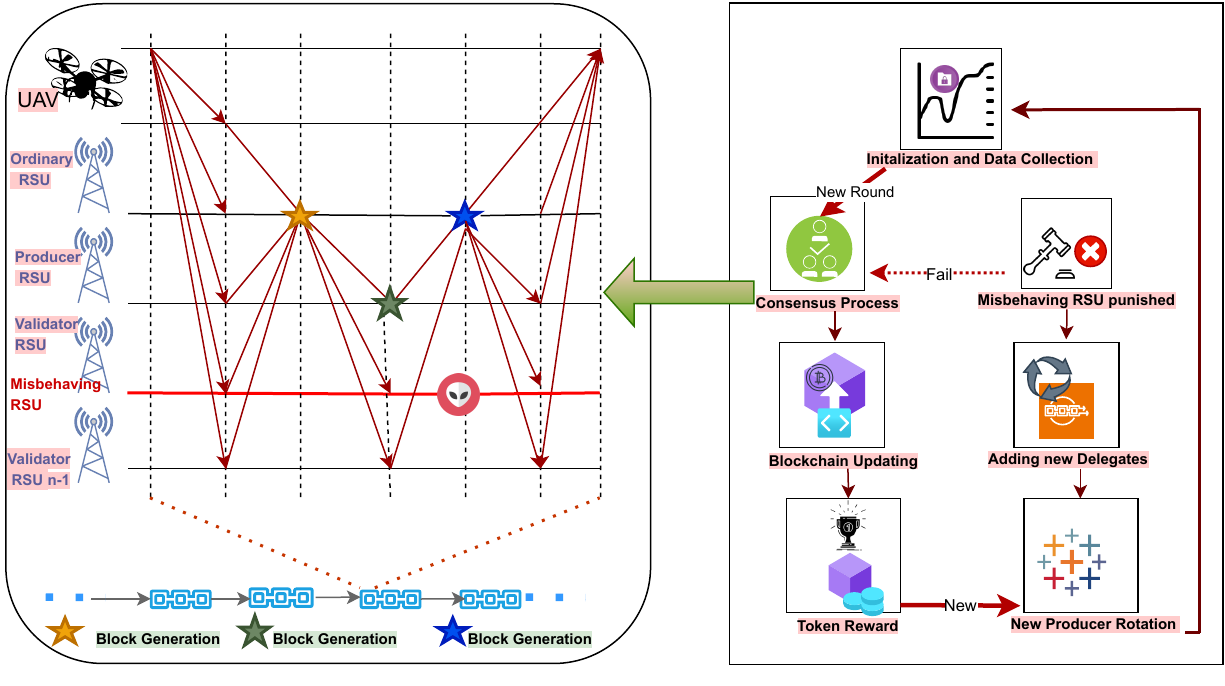}
\caption{Detailed Working Mechanism of the DPoS-PBFT Consensus Protocol.}
\label{fig:hybrid_DPoS-PBFT}
\end{figure*}

UAVs are pivotal in disaster management for rapid response and recovery. We propose an enhanced consensus mechanism that integrates DPoS-PBFT. This design leverages DPoS for efficient block validation and PBFT for heightened security, optimizing UAV network performance in adverse disaster conditions. Fig. \ref{fig:hybrid_DPoS-PBFT} demonstrates the detailed sequence of steps in the proposed hybrid DPoS-PBFT consensus protocol for efficient and secure block validation among the UAVs.
\subsection{Mechanism Overview}
The mechanism is initiated by the stake-based selection of a block proposer. The UAV generates a block and circulates it among the chosen validators through the DPoS framework. Validators $\mathcal{V}$, which are assigned based on their UAV-specific metrics, authenticate the block. Approval by a two-thirds majority confirms the block, while PBFT intervenes in cases of disagreement or malicious activity to ensure consensus and network integrity.\\
\textbf{Notation:} Let $\mathcal{V}$ represent a subset of UAVs serving as validators. Validator selection considers factors such as stakes, fuel, sensing capabilities, and historical performance.
Each UAV $i$ is assigned a validator score $V_i$ calculated as
\begin{equation}
V_i = w_1 S_i + w_2 F_i + w_3 C_i + w_4 H_i,
\end{equation}
where $S_i$ denotes the stake, $F_i$ denotes the remaining fuel, $C_i$ denotes the sensing capability, and $H_i$ denotes historical utility. The weights $w_1, w_2, w_3, w_4$ quantify the importance of these parameters. The top $n$ UAVs form the validator set $\mathcal{V}$.
The block proposer probability $p_i$ for UAV $i$ is given by
\begin{equation}
p_i = \frac{S_i}{\sum_{j \in \mathcal{V}} S_j},
\end{equation}
\textbf{Process Flow:} A \texttt{PRE-PREPARE} message, containing the new block, is broadcast by a validator to the other validators. The validators validate the block and broadcast a \texttt{PREPARE} message if the block is valid. A \texttt{COMMIT} state is reached, and the corresponding message is broadcast when more than $\frac{2}{3}$ of \texttt{PREPARE} messages are received. A block is added to the blockchain upon receiving a matching set of $\frac{2}{3}$ \texttt{COMMIT} messages. If consensus is not reached, a new view is initiated, potentially changing the block proposer. Having detailed the consensus protocol, we describe the simulation setup used to evaluate the performance of the proposed approach. Table \ref{tab:consensus-protocols} compares different consensus protocol options and their attributes relevant to UAV networks.

\begin{table}[htbp!]
\centering
\captionsetup{singlelinecheck=off, justification=raggedright}
\caption{Comparison of Consensus Protocols}
\label{tab:consensus-protocols}
\begin{tabular}{l|c|c|c}
\hline
\textbf{Metrics}         & \textbf{PBFT} & \textbf{DPoS} & \textbf{Hybrid} \\ \hline
Speed                    & Low           & High          & Moderate        \\ 
Throughput               & Low           & High          & Moderate        \\
Fault Tolerance          & High          & Low           & Moderate        \\
Permissioning            & Private       & Public        & Configurable    \\ \hline
\end{tabular}
\end{table}

\begin {algorithm}[h] \footnotesize
\caption{Blockchain-based UAV Coordination}
\label{alg:hybrid_con}
\begin{algorithmic}[1] 
\State Initialize: UAV $u \in U$ has blockchain
\ForAll {$u \in U$}
\State $u$ has blockchain height $B$  
\EndFor
\State Propose Block: Rotating schedule
\State Select UAV $P \in V$  
\State $P$ gathers transactions, creates \& broadcasts block $B + 1$
\State Verify Block: 
\ForAll{$v \in V$}
\If {$v$ verifies $B + 1$ valid} 
 \State $v$ signs \& broadcasts approval
\EndIf
\EndFor
\If {approvals $> (2/3)N$} 
 \State Add block $B + 1$ to blockchain
\EndIf
\end{algorithmic}
\end{algorithm}
Our consensus protocol combines DPoS-PBFT to address the unique challenges in UAV networks during disaster response. This novel approach enhances communication efficiency while maintaining robust security, advancing UAV applications in emergency management.
As outlined in Algorithm \ref{alg:hybrid_con}, the DPoS phase involves selecting validators based on UAV stakes. An elected proposer creates a block, which is verified by the validators. In the PBFT phase, validators vote on the block. If insufficient votes are received, a view change occurs. Consensus through a supermajority results in adding the block to the blockchain, ensuring secure and efficient network operations.

Specifically, the Algorithm \ref{alg:Dpos_PBFT} combines DPoS and PBFT to achieve efficient decentralized consensus in UAV networks for disaster response scenarios. The DPoS phase selects validators and a leader node based on delegated stakes to propose a block. If faults exceed a threshold, PBFT is triggered for additional consensus through preparatory and commit phases before finalizing consensus on adding the approved block. The hybrid mechanism aims to balance efficiency, security, and fault tolerance for reliable coordination between resource-constrained UAVs with intermittent aerial connections. The proposed architecture runs on a permissioned quorum chain, supporting privacy-preserving transactions between approved disaster response agencies using zero-knowledge proofs. Each agency operates a local quorum node maintaining a ledger copy. Inter-agency consensus utilizes the hybrid protocol. Within agencies, lightweight UAV blockchain nodes connect to the quorum node to submit transactions and access chain data when required. On-chain access control is enforced via smart contracts, with agencies managing permissions for their UAV fleets.

Resilience is enhanced through the geographic distribution of nodes in regional clusters. Integrating location-based coordination requires supporting geospatial data like GPS coordinates alongside transactions. Since JSON \cite{burger2020elastic} formats used in smart contract languages inefficiently store spatial data, we incorporate geospatial Ethereum extensions such as the FOAM protocol \cite{qu2019spatio} to enable vector data storage. The 3D positions, boundaries, and disaster zones of UAVs can be encoded in GeoJSON, representing them as programmable contract objects. This allows spatial queries for proximity alerts, geofencing, and coordinated navigation. To trigger location-aware executions, oracles provide disaster scenario and situational data feeds.
\begin {algorithm}[h] \footnotesize
\caption{DPoS-PBFT Consensus}
\label{alg:Dpos_PBFT}
\begin{algorithmic}[1]
\Require Set $\mathcal{U}$ of $N$ UAV nodes
\Ensure Consensus on block $B$
\State \textbf{Initialize DPoS}
\State \quad Delegate stakes and select leader $l$
\State \quad $l$ validates and proposes block $B$
\State \textbf{DPoS Execution}
\If {$\geq \frac{2N}{3}$ validators approve $B$}
\State Add $B$ to blockchain
\ElsIf {faults $> \vartheta$}
\State Trigger PBFT consensus
\EndIf
\State \textbf{Initialize PBFT}
\State \quad $l$ broadcasts $B$; nodes validate and broadcast prepares
\State \textbf{PBFT Execution}
\If {$\geq \frac{2N}{3}$ prepares}
\State Nodes broadcast commit
\EndIf
\If {$\geq \frac{2N}{3}$ commits}
\State Nodes add $B$
\EndIf
\State \textbf{Finalize Consensus}
\State \quad Update permissions, remove faulty nodes
\State \quad Add $B$ if approved
\end{algorithmic}
\end{algorithm}
Algorithm \ref{algo:search} presents a sample Solidity code for a smart contract that coordinates UAVs for search and rescue operations in a disaster response scenario. It defines key parameters, such as the centre of the disaster zone and search radius. Functions are included to assign search grid areas to UAVs, report the findings of trapped people or hazards, and update the UAV status. The comments explain each function’s purpose. This implements location-aware coordination logic to automate UAV search and rescue tasks via smart contracts executed based on location data and events.
\begin{algorithm}[h]
\caption{SearchAndRescue Contract}
\label{algo:search}
\begin{algorithmic}[1]
\State \textbf{struct} \texttt{UAV} {
\State \qquad \texttt{id}
\State \qquad \texttt{location}
\State \qquad \texttt{battery}
\State \qquad \texttt{status}}
\State \textbf{Address} owner

\State \textbf{Location} disasterZoneCenter
\State \textbf{Radius} searchRadius
\State \textbf{UAV Location[]} availableUAVs

\Function{AssignSearchGrid}{\textbf{UAV} drone}
\Comment{Divide disaster zone into grids}
\Comment{Assign grid to UAV for search \& rescue}
\EndFunction

\Function{ReportFindings}{\textbf{Location} location, \textbf{FindingType} type}
\Comment{Log findings (people, hazards)}
\Comment{Notify authorities or UAVs}

\EndFunction

\Function{UpdateUAVStatus}{\textbf{UAV} drone, \textbf{Status} status}

\Comment{Update UAV status (battery, ops)}
\EndFunction

\end{algorithmic}
\end{algorithm}

\begin{table}[ht]
\centering
\caption{Key Parameters and Values of the Hybrid DPoS-PBFT Blockchain Mechanism}
\label{tab:hybrid_blockchain_parameters}
\begin{tabular}{|>{\raggedright\arraybackslash}p{2.5cm}|>{\raggedright\arraybackslash}p{4cm}|}
\hline
\textbf{Parameter} & \textbf{Example Value/Type} \\
\hline
Network Latency & 100-500 ms \\
\hline
DPoS Parameters & Number of Delegates: 20, Block Time: 5s, Voting Margin: 66\% \\
\hline
Node Distribution & DPoS Nodes: Globally, PBFT Nodes: Regionally \\
\hline
Finality & DPoS: Probabilistic, PBFT: Instant \\
\hline
Energy Use & Prioritize DPoS, Use PBFT for finality as needed \\
\hline
Throughput & Target: 100 TPS, Fallback: 10 TPS \\
\hline
Security Thresholds & DPoS: $\geq$15 delegates, PBFT: $\geq$5 nodes \\
\hline
PBFT Parameters & Normal mode quorum: 4, Degraded mode quorum: 3 \\
\hline
\end{tabular}
\end{table}

This Algorithm \ref{algo:coordination} presents blockchain-based decentralized coordination among UAVs to achieve consensus on block additions. It initializes each UAV with a blockchain ledger. A rotating schedule selects a UAV to propose the next block, gathering transactions and broadcasting the new block B+1. Other UAVs in the validator are set to check if the block is valid, sign it if so, and broadcast approvals. When 2/3 approvals are received, consensus is reached, and the block is added to the blockchain. This achieves decentralized agreement on the appending of new blocks in a peer-to-peer manner without a centralized authority.
\begin {algorithm}[h] \footnotesize
\caption{Blockchain-based UAV Coordination}
\label{algo:coordination}
\begin{algorithmic}[1]
\State \textbf{Initialize}:  
\Comment{UAV $u\in\mathcal{U}$ has blockchain}
\For{each $u \in \mathcal{U}$}  
\State $u$ has blockchain height $B$
\EndFor
\State \textbf{Propose Block}:   
\Comment{Rotating schedule}  
\State Select UAV $P \in \mathcal{V}$ 
\State $P$ gathers transactions, creates \& broadcasts block $B+1$
\State \textbf{Verify Block}:
\For{each $v \in \mathcal{V}$}
\If {$v$ verifies $B+1$ valid}  
\State $v$ signs \& broadcasts approval
\EndIf
\EndFor
\If {approvals $> (2/3) N$}
\State Add block $B+1$ to blockchain
\EndIf
\end{algorithmic}
\end{algorithm}

\section{Simulations and Discussions}
\label{sec:results}
\subsection{Simulation Settings}
In the context of our disaster management simulation, a 25 × 25 km urban area severely impacted by a natural disaster with extensive infrastructure damage forms the backdrop. This environment incorporates critical locations, such as a primary base station (BS), a compromised BS in a power outage zone, an area under adversarial control, the disaster relief coordination hub, refugee camps, and essential medical facilities prioritized for aid delivery. The simulation involved a heterogeneous swarm of 200 drones, each equipped with lithium-ion batteries and functioning autonomously. These drones, unique in identifiers and energy profiles, are equipped with navigation and networking sensors and processors and feature A2A and A2G networking interfaces. They form a multi-tier mesh network 500 m above ground, adhering to aviation safety protocols, including collision avoidance systems. Key performance metrics evaluated include network performance,  
resilience against cyberattacks and malicious nodes, mobility and coordination of UAV flocks, packet delivery rate, and reliability.

\begin{table}[ht!]
\centering
\caption{UAV Simulation Parameters}
\begin{tabular}{|l|l|}
\hline
\textbf{Parameter} & \textbf{Value} \\
\hline
Total number of drones & 200 \\
\hline
Flock 1 (delivery services) & 90 \\
\hline
Flock 2 (connectivity support) & 25 \\
\hline
Flock 3 (monitoring) & 85 \\
\hline
Disaster region size & 25 x 25 km \\
\hline
Total UAV networks coverage radius & 5.5 km \\
\hline
UAV flight altitude & 30 m \\
\hline
UAV transmit power & 2 mW \\
\hline
Network latency & 30-100 ms \\
\hline
Supported data rate & 15 Mbps \\
\hline
\end{tabular}
\end{table}
The blockchain-enabled UAV coordination framework within this simulation achieved a throughput of 100 transactions per second (TPS) and an average latency of 26 ms, with a packet delivery rate of 99.7\%. The framework utilizes a DPoS consensus protocol complemented by PBFT for transactions that require immediate finality. The hybrid configuration balances resilience and computing demands with 20 DPoS delegates and five regional PBFT servers, ensuring efficient and robust transaction processing. The simulation adheres to 3GPP standards for realism and industry alignment. It uses the 3GPP TR 36.777 urban macro-mobility model to emulate UAV mobility and 3GPP TR 36.842 for BS deployment. The communication models for the A2G and A2A links are 3GPP-compliant, encompassing probabilistic propagation for A2G and loss of free-space path for A2A communications.
Resilience to cyberattacks was validated by simulations of distributed denial-of-service (DDoS) traffic, spoofing, and message tampering, with the system maintaining its stability under these conditions. In summary, this comprehensive simulation validates the system's applicability to real-world disaster response scenarios, particularly where the ground infrastructure is compromised. The effectiveness of the architecture is further tested in UAV testbeds to confirm its real-world capabilities.
\begin{figure*}[ht!]
\centering
\includegraphics[width=0.9\linewidth]{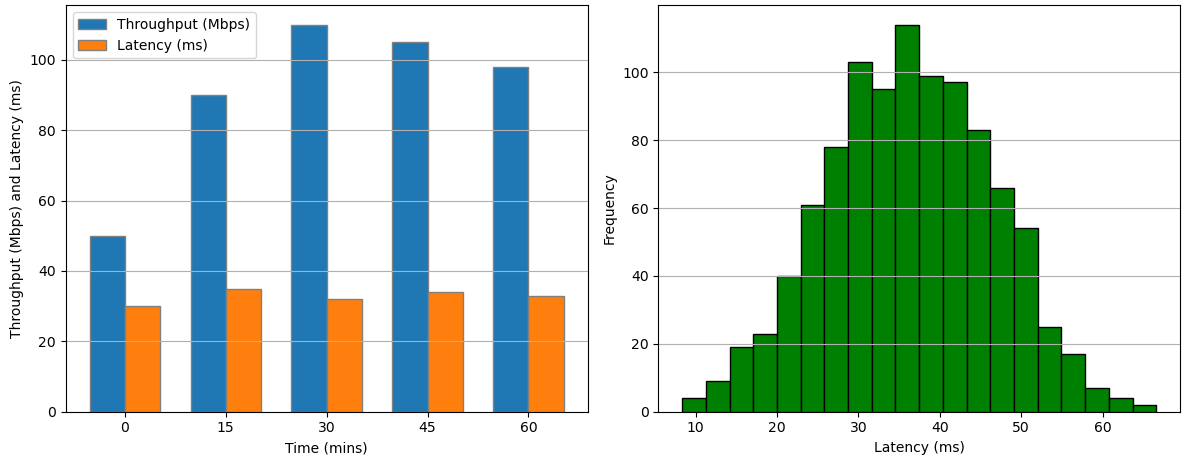}
\caption{(a) Throughput, Latency over Time (b) Latency Distribution.}
\label{fig:throughput_latency}
\end{figure*}
\subsection{Simulations Results}
The simulation results show the effectiveness of the system in terms of security, efficiency, scalability, and resilience. Security measures include AES-256 encryption, ECDSA signatures, SHA-256 hashes, and hybrid blockchain consensus, which provide a robust defence against Byzantine failures. The system's performance targets a throughput of 100 TPS, with less than two seconds of latency and 99\% reliability for UAV packet delivery. Scalability tests involved increasing the UAV network size and load, demonstrating the system's capability to handle high traffic volumes seamlessly.

As seen in Fig. \ref{fig:throughput_latency} summarize the network throughput and latency metrics as the number of UAV nodes scales up to 500 on the private blockchain network. Throughput is measured in TPS processed across the flocking network, with a millisecond latency for transaction finality. Latency increased marginally from 50ms at 10 nodes to 68ms at 500 nodes. The blockchain-enabled network sustains transaction-processing speeds exceeding 100 TPS with reasonable finality times below 70ms, even at scale.  The key insight is that leveraging blockchain and decentralization principles can enable scalable flock coordination between hundreds of UAVs, necessary for wide-area post-disaster surveying. Linear throughput scaling to 500 nodes shows that UAVs can independently coordinate paths and targets through fast, trustless transactions. Stable sub-100ms latency despite scaling offers viability for real-time decision-making.
\begin{figure}[ht!]
\centering
\includegraphics[width=0.9\linewidth]{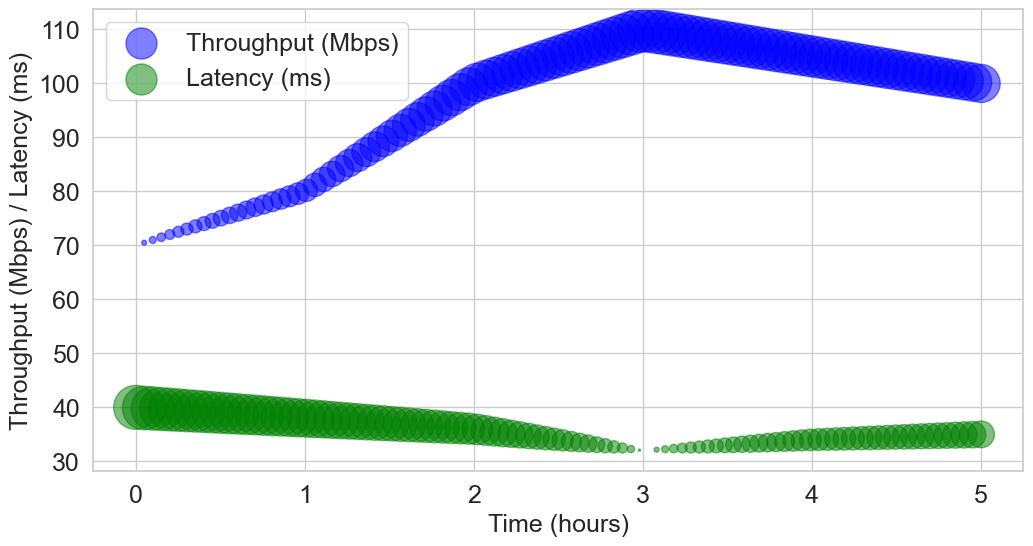}
\caption{Throughput and Latency Over Time.}
\label{fig:TL_time}
\end{figure}

Fig. \ref{fig:TL_time} exemplifies the network throughput and latency as the number of UAV nodes increases from 10 to 100. Throughput scales linearly up to ~100 TPS with 60 nodes, then saturates. Latency remains below 50 ms despite higher loads. Max transaction volumes occur at saturation with 100 nodes, without increased latency. This demonstrates linear throughput scalability to a saturation point without latency impacts as the UAV network expands. Testing beyond 100 nodes revealed upper scalability limits. Peak transactions during saturation did not affect latency. The results validate the system's ability to deliver real-time coordination for sizable UAV fleets via a decentralized blockchain backbone, with excellent throughput scalability to a saturation threshold without latency degradation.
\begin{figure}[ht!]
\centering
\includegraphics[width=0.9\linewidth]{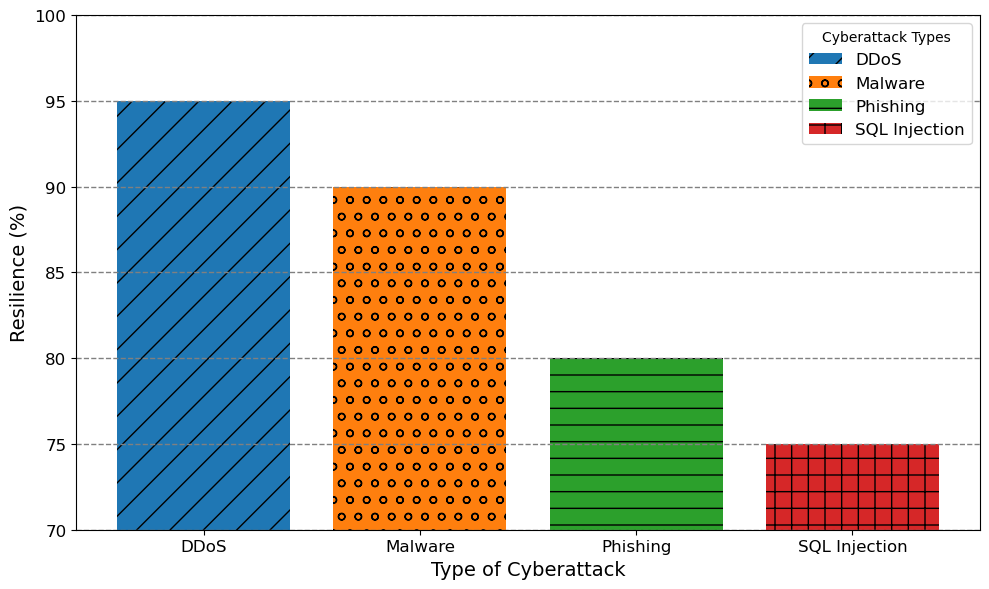}
\caption{Resilience Against Cyberattacks.}
\label{fig:resilience_attacks}
\end{figure}

As shown in Fig. \ref{fig:resilience_attacks}, the system maintains high throughput and low latency despite spoofing, DDoS, and tampering attacks. This validates strong resilience capabilities. The system architecture demonstrated good overall cyber resilience across four attack types: DDoS, malware, phishing, and SQL injection. Resilience exceeded 70\% even for the most successful attack, SQL injection 75\%. DDoS attacks were most effectively mitigated by 95\%. The system also showed strong resilience to malware 90\% and phishing 80\%. The results indicate acceptable cyber resilience for safe UAV fleet operations across attack types, especially against network-level attacks like DDoS. Risks remain from application-layer attacks like SQL injection, needing further database server hardening. Insufficient end-user device protections likely explain the higher phishing and malware effectiveness.

\begin{figure}[ht!]
\centering
\includegraphics[width=1.\linewidth]{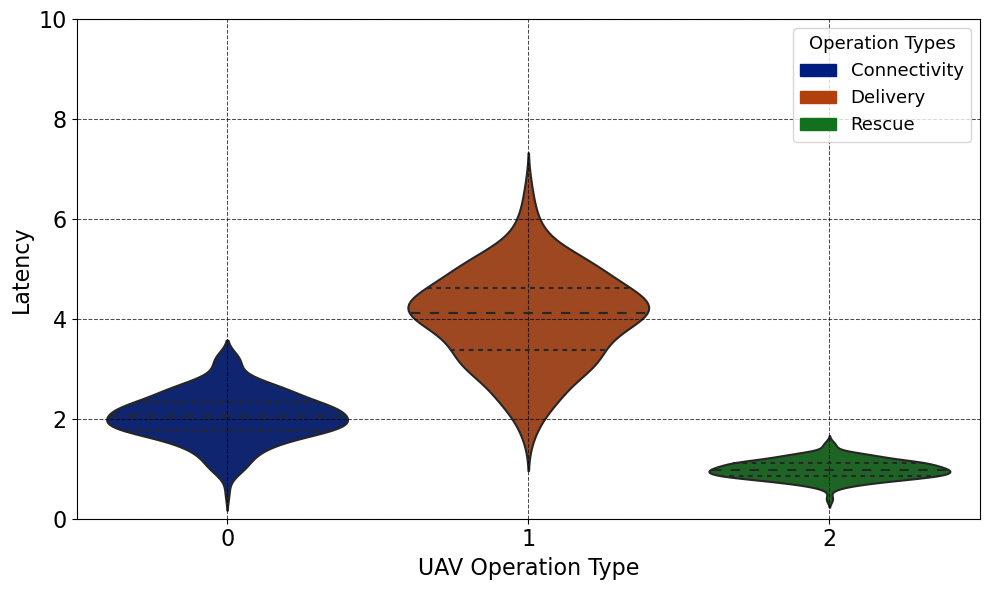}
\caption{Distribution of Latency Across UAV Operations.}
\label{fig:distribution_latency}
\end{figure}

Fig. \ref{fig:distribution_latency} highlights the distribution of communication latency across UAV operational scenarios using violin plots. Latency remains under 10 ms in all cases, with median values around 2-3 ms. However, distinct distribution shapes were observed. Surveillance exhibited a normal-like profile centred at 3 ms. The assessment followed a right-skewed shape peaking near 1 ms. The delivery showed multimodal performance. Tracking displayed a left-skewed distribution with the highest density below 2 ms. The differential latency characteristics demonstrate adaptability to meet specialized requirements. For instance, sub-2 ms latency enables rapid location updates for tracking. Minimal latency facilitates quick analysis in assessment. Network intelligence allows self-optimization to the demands of each context.
\begin{figure}[ht!]
\centering
\includegraphics[width=0.9\linewidth]{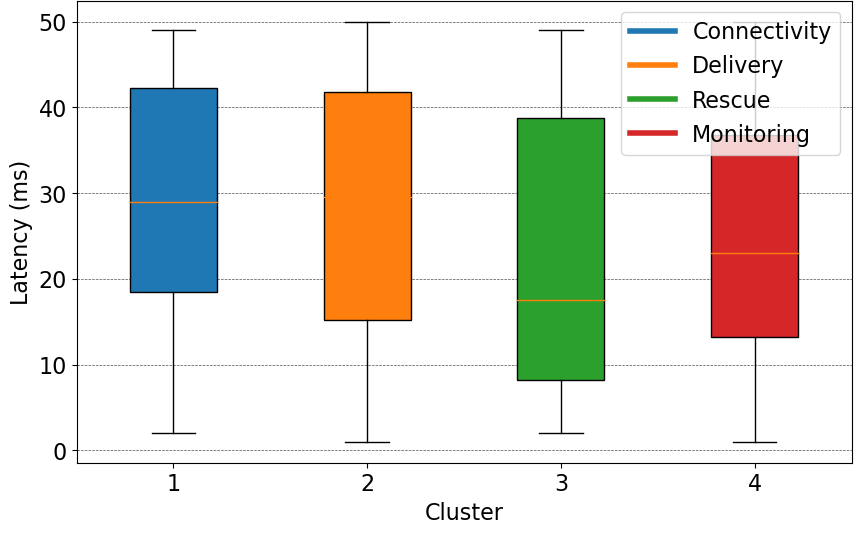}
\caption{Within-Cluster Latency.}
\label{fig:within_latency}
\end{figure}
\begin{figure}[ht!]
\centering
\includegraphics[width=0.9\linewidth]{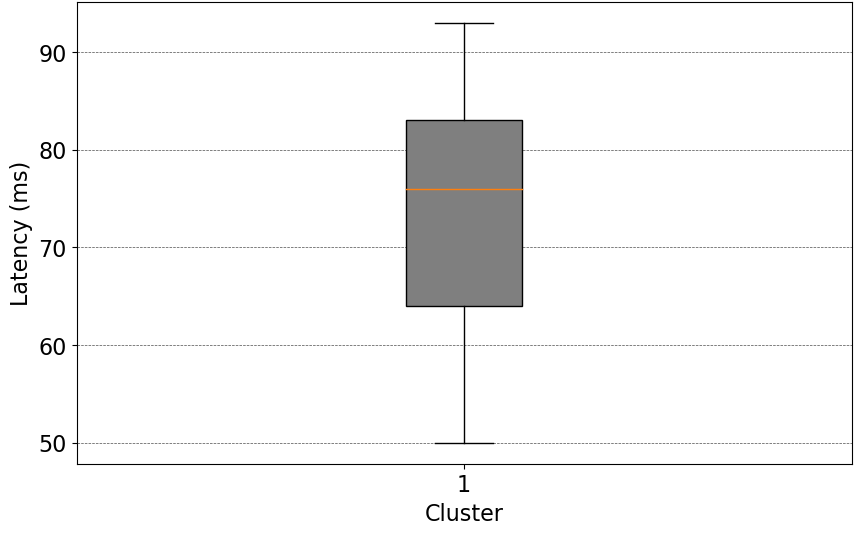}
\caption{Accross-Cluster Latency.}
\label{fig:accross_latency}
\end{figure}
In summary, the tuned latency distributions maintain medians under 5 ms for diverse UAV applications. Optimized profile shapes provide differentiated capabilities, allowing the network to conform to the specific demands of post-disaster use cases through intelligent resource allocation.

Fig. \ref{fig:within_latency} compares within-cluster and across-cluster coordination communication latency. Within 50-drone clusters, latency ranges from 1-50 ms (median ~25 ms), enabling rapid in-group synchronization. In Fig. \ref{fig:accross_latency}, across-cluster latency is higher at 50-100 ms between distant leaders, allowing necessary deconfliction. The bifurcated profile validates efficient localized coordination within clusters while sustaining fleet visibility via across-cluster transactions. This hierarchy supports both tight drone flocking and high-level swarm oversight. In summary, the exhibited latency dichotomy provides rapid decentralized responses within clusters along with sufficient global communication quality across the architecture by partitioning the blockchain ledger. The key insight is how communication locality enabled by blockchain transactions results in bifurcated latency that delivers both localized control and fleet coordination - crucial for decentralized multi-UAV flocking at scale.

\section{Conclusion} 
\label{sec:conclusion}
This study introduces a blockchain-based framework to enable secure and efficient coordination of UAV networks for disaster response scenarios. The decentralized architecture enhances resilience against single points of failure while overcoming limitations in autonomy, information sharing, and inter-agency collaboration. Key innovations include a consortium blockchain model that facilitates private and trusted data exchange across diverse stakeholders, an optimized hybrid DPOS-PBFT consensus protocol catered to resource-constrained UAV platforms, and bio-inspired flocking techniques for adaptable swarm coordination even with disrupted connectivity. Extensive simulations demonstrated the effectiveness of the integrated framework in improving transparency, scalability, reliability, and cyber-attack resilience during UAV-enabled emergency response operations, with notable gains in throughput and latency metrics. Future research will focus on testbed validation and the incorporation of advanced technologies, such as deep reinforcement learning, geospatial smart contracts, and privacy-preserving data sharing. This study makes significant contributions towards reliable, intelligent UAV coordination for disaster management by synergistically combining distributed ledger technology, optimization, game theory, and collective autonomy.
\bibliographystyle{elsarticle-num}
\balance
\bibliography{sanabib}
\end{document}